\documentclass[aps,prl,superscriptaddress,twocolumn,showpacs,amsmath,amssymb,floats]{revtex4}

\usepackage{txfonts}
\usepackage{amssymb}
\usepackage{graphicx}

\begin{document}
	
\title{Orbital-dependent modulation of the superconducting gap in uniaxially strained Ba$_{0.6}$K$_{0.4}$Fe$_2$As$_2$}

\author{L. Chen}
\affiliation{International Center for Quantum Materials, School of Physics, Peking University, Beijing 100871, China}

\author{T. T. Han}
\affiliation{International Center for Quantum Materials, School of Physics, Peking University, Beijing 100871, China}

\author{C. Cai}
\affiliation{International Center for Quantum Materials, School of Physics, Peking University, Beijing 100871, China}

\author{Z. G. Wang}
\affiliation{International Center for Quantum Materials, School of Physics, Peking University, Beijing 100871, China}

\author{Y. D. Wang}
\affiliation{International Center for Quantum Materials, School of Physics, Peking University, Beijing 100871, China}

\author{Z. M. Xin}
\affiliation{International Center for Quantum Materials, School of Physics, Peking University, Beijing 100871, China}

\author{Y. Zhang}\email{yzhang85@pku.edu.cn}
\affiliation{International Center for Quantum Materials, School of Physics, Peking University, Beijing 100871, China}
\affiliation{Collaborative Innovation Center of Quantum Matter, Beijing 100871, China}

\date{\today}

\begin{abstract}

Pairing symmetry which characterizes the superconducting pairing mechanism is normally determined by measuring the superconducting gap structure ($|\Delta_k|$). Here, we report the measurement of a strain-induced gap modulation ($\partial|\Delta_k|$) in uniaxially strained Ba$_{0.6}$K$_{0.4}$Fe$_2$As$_2$ utilizing angle-resolved photoemission spectroscopy and $in$-$situ$ strain-tuning. We found that the uniaxial strain drives Ba$_{0.6}$K$_{0.4}$Fe$_2$As$_2$ into a nematic superconducting state which breaks the four-fold rotational symmetry of the superconducting pairing. The superconducting gap increases on the $d_{yz}$ electron and hole pockets while it decreases on the $d_{xz}$ counterparts. Such orbital selectivity indicates that orbital-selective pairing exists intrinsically in non-nematic iron-based superconductors. The $d_{xz}$ and $d_{yz}$ pairing channels are balanced originally in the pristine superconducting state, but become imbalanced under uniaxial strain. Our results highlight the important role of intra-orbital scattering in mediating the superconducting pairing in iron-based superconductors. It also highlights the measurement of $\partial|\Delta_k|$ as an effective way to characterize the superconducting pairing from a perturbation perspective.

%($\Delta_{xz}$ and $\Delta_{yz}$)
\end{abstract}


\maketitle

Pairing symmetry describes the symmetry of pairing interactions and is hence an important characteristic of a superconductor. To determine the pairing symmetry, measuring the superconducting gap distribution in momentum space ($|\Delta_k|$) is an effective way. For example, the superconducting gap is isotropic in momentum space for a s-wave superconductor, while for a d-wave superconductor, like cuprates, the superconducting gap is anisotropic with gap nodes along the diagonal direction of Brillouin zone. For iron-based superconductors (IBSs), however, due to the complexity of their multi-orbital and multi-band nature, the pairing symmetry of IBSs is still under hot debates \cite{ref1,ref2,ref3}. Theoretically, different pairing models predict different pairing symmetries, including the s$_{++}$-wave that is mediated by orbital-fluctuation \cite{ref4}, the s$_\pm$-wave that is mediated by the ($\pi$, 0) spin-fluctuation or next-nearest neighbor exchange interaction \cite{ref5,ref6,ref7}, the d-wave that is mediated by the ($\pi$, $\pi$) spin fluctuation or nearest-neighbor exchange interaction, etc \cite{ref8,dwave}. Depending on the detailed Fermi surface topologies and band symmetries, different pairing symmetries could further mix with each others, making the determination of pairing symmetry in IBSs very difficult. \cite{ref9,ref10,ref39}.

Experimentally, the gap structure of IBSs has been studied extensively by angle-resolved photoemission spectroscopy (ARPES) and other techniques. While most IBSs show multi-gap behaviors with isotropic gap structures \cite{ref1,ref11,ref38,ref12,ref13}, strong gap anisotropy and gap nodes were observed in LiFeAs, FeSe, BaFe$_2$(As$_{1-x}$P$_x$)$_2$, KFe$_2$As$_2$, etc \cite{refLiFeAs1,refLiFeAs2,ref15,ref16,ref17,ref18}. The observed gap structures are diverse and material-dependent, making the measurements of $|\Delta_k|$ insufficient to determine the pairing symmetry of IBSs. Therefore, instead of studying the pristine superconducting phase, we focus on the perturbed superconducting phase where crucial clues may be found by studying how the superconducting gap responds to a perturbation field. According to previous studies, various properties of IBSs, including resistivity \cite{ref20, strainreview}, electronic structure \cite{strainreview,ref21,ref22}, magnetic excitation \cite{ref23}, etc., response strongly to the uniaxial strain. The system breaks the four-fold rotational symmetry under uniaxial strain due to the presence of strong nematic fluctuation.

Here, utilizing ARPES and a piezo-based $in$-$situ$ strain-tuning device, we show that the uniaxial strain also breaks the four-fold rotational symmetry of $|\Delta_k|$ and drives an optimally doped iron-arsenide superconductor Ba$_{0.6}$K$_{0.4}$Fe$_2$As$_2$ into a nematic superconducting phase. We find that the strain-induced gap modulation ($\partial|\Delta_k|$) is strongly orbital-selective. Such a gap modulation cannot be explained by a mixing of s- and d-wave pairing symmetries, but instead, can be well explained by a strain-induced imbalance of the $d_{xz}$ and $d_{yz}$ pairing channels. Our results thus show that the orbital-selective pairing exists intrinsically in non-nematic IBSs. The orbital selectivity can be further tuned by the uniaxial strain, which provides an effective experimental method for determining the pairing symmetry of IBSs from a perturbation perspective.

\begin{figure}[t]
	\includegraphics[width=8.7cm]{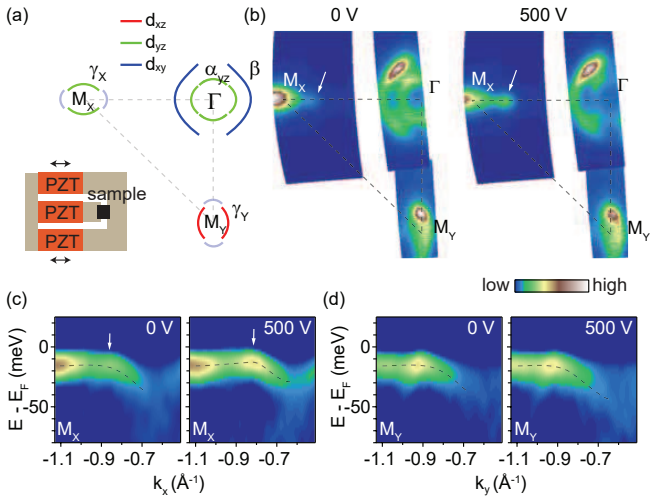}
	\caption{(color online) (a) Schematic drawing of the observable Fermi surface sheets of Ba$_{0.6}$K$_{0.4}$Fe$_2$As$_2$ \cite{ref26,ref27}. The inset panel illustrates the working principle of our $in$-$situ$ strain-tuning device \cite{ref21,ref26}. (b) Fermi surface mapping taken at 0~V and 500~V. The white arrows highlight the change of photoemission intensity under uniaxial strain. (c) Second derivative images of the energy-momentum cuts taken along the $\Gamma$-M$_X$ direction at 0~V and 500~V. Black dashed lines guide the eye to the band dispersions. (d) is the same as (c) but taken along $\Gamma $-M$_Y$ direction. The raw data can be found in the supplemental material \cite{ref26}. }\label{fig1}

\end{figure}

High-quality single crystals of Ba$_{0.6}$K$_{0.4}$Fe$_2$As$_2$ were synthesized using a self-flux method with a nominal composition of Ba~:~K~:~Fe~:~As~=~0.5~:~0.5~:~4~:~4 \cite{ref25}. The superconducting transition temperature (T$_c$) is around 38~K as confirmed by the magnetic susceptibility measurement. The chemical composition is determined by the energy disperse spectroscopy. ARPES measurements were performed at Peking University using a DA30L analyser and a helium discharging lamp. The photon energy is 21.2~eV. The overall energy resolution is around 6~meV and the angular resolution is around 0.3$ ^\circ $. The samples were cleaved $in$-$situ$ and measured in ultrahigh vacuum with a base pressure better than $6\times10 ^{-11}$~mbar. All data were taken at 20~K. Our home-made piezo-based strain-tuning device is shown in Fig.~\ref{fig1}(a), whose details can be found in Ref. \cite{ref21,ref26,sup1}. The uniaxial strain is applied along the $x$ direction. The strain strength is  characterized by microscope and X-ray diffraction \cite{ref21,ref26} and is represented by the piezo voltage. At 500~V, around 0.5\% tensile strain could be achieved.

\begin{figure}[t]
	\includegraphics[width=8.7cm]{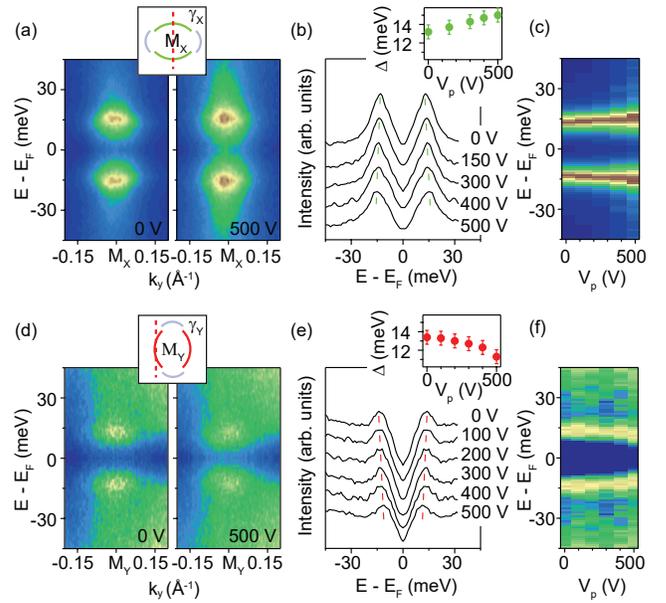}
	\caption{(a) Symmetrized energy-momentum cuts taken across the $ \gamma_X $ electron pocket at 0~V and 500~V. (b) Strain dependence of the symmetrized energy distribution curves (EDCs) taken at the Fermi crossing (k$ _F $) of the $\gamma_X$ electron pocket. The k$ _F $ is determined as the momentum where the gap reaches a minimum. The inset panel shows the strain dependence of the superconducting gap. The gap magnitudes were determined using the gap fitting process \cite{ref36}. The error bar is estimated by considering the fitting error, the energy resolution, and the error of $k_F$. (c) is a merged image of the symmetrized EDCs shown in (b). (d), (e), and (f) are the same as (a), (b) and (c) but taken on the $\gamma_Y$ electron pocket. }\label{fig2}
\end{figure}

According to previous studies \cite{ref26,ref27,ref30,ref31}, the Fermi surface of Ba$_{0.6}$K$_{0.4}$Fe$_2$As$_2$ consists of three hole pockets at the Brillouin zone center ($\Gamma$) and two electron pockets at the Brillouin zone corner (M) [Figs.~\ref{fig1}(a) and \ref{fig1}(b)]. For the $d _{xz}$/$d_{yz}$ inner hole pockets, we only observe the $d_{yz}$ sections ($\alpha_{yz}$) due to the orbital-selectivity of our polarized light \cite{ref26,ref33,ref34}. For the electron pockets, due to the glide-mirror-symmetry of the iron-arsenide plane, the unfolded bands in the one-iron Brillouin zone show prominent photoemission intensity \cite{ref26,ref35}. As a result, the $d_{yz}$/$d_{xy}$ and $d_{xz}$/$d_{xy}$ electron pockets ($\gamma_X$ and $\gamma_Y$) are observed at the M$_X$ and M$_Y$ points, respectively.

When applying a tensile strain on the sample, the response of Fermi surface is obvious, especially around the M points [Fig.~\ref{fig1}(b)]. Photoemission intensity increases around the M$_X$ point while decreases around the M$_Y$ point. According to the previous ARPES studies \cite{strainreview,ref21,ref22}, the electron bands around the M points are constructed by the $d_{xz}$ and $d_{yz}$ orbitals and therefore are sensitive to the uniaxial strain. It is expected that when the uniaxial strain breaks the four-fold rotational symmetry of the system, the $d_{xz}$ and $d_{yz}$ bands would shift in two opposite directions. Here, as shown in Figs.~\ref{fig1}(c) and \ref{fig1}(d), the band dispersion along the $\Gamma$-M$_X$ direction shifts up towards the Fermi energy, while the band dispersion along the $\Gamma$-M$_Y$ direction shifts down slightly towards higher binding energy. Such anisotropy reflects a breaking of four-fold rotational symmetry, which is further confirmed by the transport measurement that shows an emergence of resistivity anisotropy in Ba$_{0.6}$K$_{0.4}$Fe$_2$As$_2$ under uniaxial strain \cite{ref26}. Note that, there is no nematic phase in the pristine Ba$_{0.6}$K$_{0.4}$Fe$_2$As$_2$ \cite{ref25}, which is also confirmed by our ARPES and resistivity data. Therefore, the symmetry breaking of electronic structure and resistivity anisotropy observed here cannot be explained as a detwinning effect. Instead, it indicates that the uniaxial strain drives Ba$_{0.6}$K$_{0.4}$Fe$_2$As$_2$ into a nematic state which breaks the four-fold rotational symmetry.

Next, we turn to the strain dependence of the superconducting gap. We first compare the superconducting gaps taken on the $\gamma_X$ and $\gamma_Y$ electron pockets. As shown in Fig.~\ref{fig2}, the superconducting gaps on the $\gamma_X$ and $\gamma_Y$ electron pockets are equal in the pristine sample as constrained by the four-fold rotational symmetry. When applying a tensile strain along the $x$ direction, the superconducting gap increases on the $\gamma_X$ pocket while decreases on the $\gamma_Y$ pocket. The strain-induced gap modulation reaches $\sim$2~meV at 500~V. Note that, we did not observe any $T_c$ suppression in uniaxially strained Ba$_{0.6}$K$_{0.4}$Fe$_2$As$_2$ \cite{ref26}. The superconductivity of our sample is also clearly characterized by the band-bending behavior and sharp superconducting coherent peaks, suggesting that the energy gap observed here is a superconducting gap. Furthermore, the validity of the strain-dependent data is confirmed by the strain-cycle and sample-crack experiments \cite{ref26}. Therefore, the anisotropic gap modulation observed on the $\gamma_X$ and $\gamma_Y$ electron pockets is a direct consequence of the uniaxial strain. Such gap anisotropy reflects a four-fold rotational symmetry breaking of the superconducting pairing, which indicates an emergence of a nematic superconducting state in uniaxially strained Ba$_{0.6}$K$_{0.4}$Fe$_2$As$_2$.

\begin{figure}[t]
	\includegraphics[width=8.7cm]{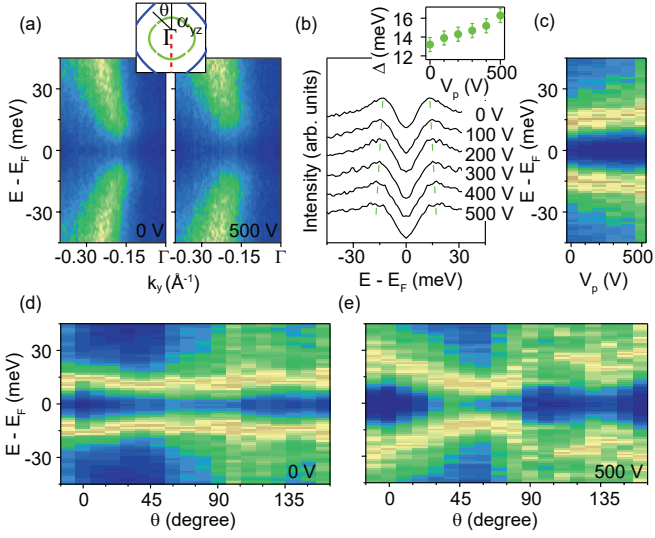}
	\caption{(a) Symmetrized energy-momentum cuts taken across the $ \alpha_{yz} $ hole pocket at 0~V and 500~V. (b) Strain dependence of the symmetrized EDCs taken at the k$ _F $ of the $ \alpha_{yz} $ hole pocket. The inset panel shows the strain dependence of the superconducting gap. (c) is a merged image of the symmetrized EDCs shown in (b). (d) Angular dependence of the superconducting gap taken on the $ \alpha_{yz} $ hole pocket at 0~V. The symmetrized EDCs taken along the $\alpha_{yz}$ hole pocket are merged into one image to better illustrate the angular distribution of the superconducting gap \cite{ref26}. (e) is the same as (d), but taken at 500~V. }\label{fig3}
\end{figure}

We then measured the strain dependence of the superconducting gap on the inner hole pocket $\alpha_{yz}$. As shown in Figs.~\ref{fig3}(a)-\ref{fig3}(c), the superconducting gap increases with the increment of tensile strain with a total increment of $\sim$3~meV at 500~V. Figs.~\ref{fig3}(d) and \ref{fig3}(e) further show the angular distributions of the  superconducting gap measured in the pristine and uniaxially strained samples. The superconducting gap is less strain-sensitive along the 45$^\circ$ and 135$^\circ$ directions,  which is expected because these directions are diagonal to the strain direction. Along the 0$^\circ$ and 90$^\circ$ directions, intuitively, the superconducting gap should increase along one direction while decrease along the other. However, as shown in Figs.~\ref{fig3}(d) and \ref{fig3}(e), the superconducting gap increases along both the 0$^\circ$ and 90$^\circ$ directions under uniaxial strain.

\begin{figure}[t]
	\includegraphics[width=8.7cm]{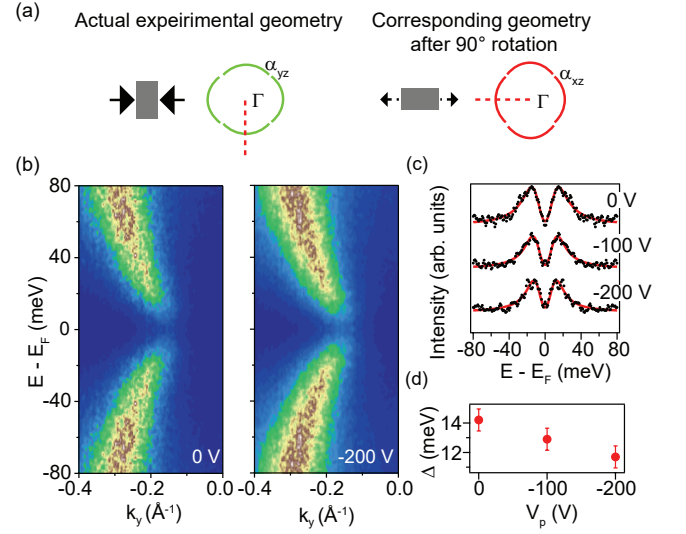}
	\caption{(a) Schematic drawing of the actual experimental geometry and corresponding geometry after 90$^\circ$ rotation. (b) Symmetrized energy-momentum cuts taken at 0~V and -200~V. (c) Strain dependence of the symmetrized EDCs taken at the $k_F$ of the inner hole pocket. The raw and fitted data are illustrated by the black dots and red lines, respectively. (d) Strain dependence of the superconducting gap on the inner hole pocket. }\label{fig4}
\end{figure}

To explain this counter-intuitive result, we need to consider the multi-orbital character of the inner hole pockets. We can only detect the $d_{yz}$ sections of the inner hole pockets due to the orbital selectivity of our polarized light, \cite{ref26,ref33,ref34}. To measure the $d_{xz}$ counterparts, we need to rotate either the beam polarization or the sample itself, both of which cannot be done in our ARPES system. Therefore, we made an equivalent measurement of the $d_{xz}$ orbital by applying a compressive strain. The trick is that the nematic domain rotates 90$^\circ$ when the strain direction changes from stretched to compressed. As a result, the $d_{yz}$ orbital in the compressed sample is equivalent to the $d_{xz}$ orbital in the stretched sample after rotating the coordinates by 90$^\circ$ [Fig.\ref{fig4}(a)]. In the actual experiment, the sample is compressed along the $x$ direction by applying a negative voltage on the peizo stacks. We measured the gap evolution of the $d_{yz}$ band in the compressed sample, which could be viewed as an equivalent measurement of the gap evolution of the $d_{xz}$ band in the stretched sample. The result shows that the superconducting gap decreases on the $d_{xz}$ sections of the inner hole pockets  (Fig.\ref{fig4}).

\begin{figure}[t]
	\includegraphics[width=6cm]{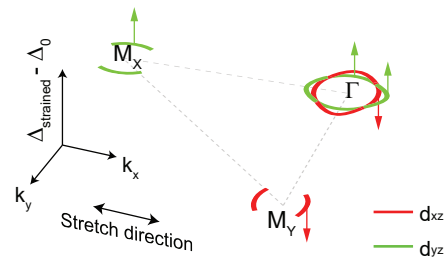}
	\caption{Schematic drawing of the gap modulation on the $d_{xz} $/$d_{yz} $ Fermi surface sheets in uniaxially strained Ba$_{0.6}$K$_{0.4}$Fe$_2$As$_2$. }\label{fig5}
\end{figure}

Figure~\ref{fig5} summarizes the strain-induced gap modulation in uniaxially strained Ba$_{0.6}$K$_{0.4}$Fe$_2$As$_2$. Here, we only focus on the $d_{xz} $/$d_{yz}$ Fermi pockets, because the superconducting gap on the $d_{xy} $ outer hole pocket is too small to be resolved due to our high experimental temperature. First, we find that the rotational symmetry breaking is represented not by an angular anisotropy, but by an orbital selectivity. Under uniaxial strain, the superconducting gap increases on the $\alpha_{yz}$ and $\gamma_X$ pockets which are mostly constructed by the $d_{yz}$ orbital, while the gap decreases on the $\alpha_{xz}$ and $\gamma_Y$ pockets which are mostly constructed by the $d_{xz}$ orbital. Second, the magnitudes of the gap modulations at $\Gamma$ and M are comparable. The strain-induced gap modulation is around 2-3~meV on both the hole and electron pockets.

To understand the strain-induced gap modulation, we consider two different scenarios. First, the strain-induced gap modulation is a small perturbation that is added to the original superconducting order parameter. Here, the original pairing symmetry of Ba$_{0.6}$K$_{0.4}$Fe$_2$As$_2$ is commonly believed to be four-fold symmetric \cite{ref1,ref11,ref38}, which suggests that the strain-induced perturbation must be rotational-symmetry-broken. This constraint excludes the four-fold symmetric pairing symmetries, such as s-wave and extended s-wave, as a cause of the strain-induced gap modulation, while indicates a possible existence of s~+~d pairing symmetry \cite{ref10,ref39}. The d-wave pairing symmetry shows opposite signs at the M$_X$ and M$_Y$ points. By mixing a d-wave pairing symmetry, the superconducting gap increases at one M point but decreases at the other, which is consistent with the strain-induced gap anisotropy observed on the $\gamma_X$ and $\gamma_Y$ pockets. However, the d-wave gap function [$\Delta$($\cos k_x-\cos k_y$)] decreases rapidly when moving from M to $\Gamma$. This contradicts to our observation that the gap modulations at $\Gamma$ and M are comparable. Furthermore, the s~+~d pairing symmetry cannot explain why the gap increases along both the 0$^\circ$ and 90$^\circ$ directions on the $\alpha_{yz}$ pocket.

These contradictions lead to the proposing of the second scenario. Instead of adding a small symmetry-broken perturbation, the uniaxial strain breaks the balance of the $d_{xz}$ and $d_{yz}$ pairing channels that exist originally in Ba$_{0.6}$K$_{0.4}$Fe$_2$As$_2$. Considering only the $d_{xz}$ and $d_{yz}$ orbitals for simplicity, the simplest description of an orbital-selective gap function would be $\Delta_{xz}$+$\Delta_{yz}$, where $\Delta_{xz}$ and $\Delta_{yz}$ reflect the pairing strength of the $d_{xz}$ and $d_{yz}$ pairing channels. $\Delta_{xz}$ and $\Delta_{yz}$ are balanced in the pristine sample as constrained by the four-fold rotational symmetry, but become imbalanced under uniaxial strain ($\Delta_{xz}$~$\neq$~$\Delta_{yz}$). This well explains the orbital-selective gap modulation observed here in uniaxially strained Ba$_{0.6}$K$_{0.4}$Fe$_2$As$_2$.

The orbital-selective superconducting pairing has been observed in FeSe \cite{FeSeSTM,ref40, ref41,ref42} which is naturally a nematic superconductor. It was found that the superconducting gap of FeSe is large on the $d_{yz}$ Fermi surface, but shows nodes on the $d_{xz}$ Fermi surface \cite{FeSeSTM,ref40}. Here, we drive Ba$_{0.6}$K$_{0.4}$Fe$_2$As$_2$ continuously from a normal superconductor to a nematic superconductor. One the one hand, our results build a connection between FeSe and iron-arsenide superconductors. Our observation of the gap increment on the $d_{yz}$ Fermi surface is consistent with the large gap observed on the $d_{yz}$ Fermi surface of FeSe, showing a dominating role of the $d_{yz}$ orbital in the nematic superconducting state. On the other hand, we show that the orbital-selective pairing is a universal behavior that exists intrinsically in the non-nematic iron-based superconductors. Such orbital-selective pairing originates from an intra-orbital pair-scattering of electrons which is most likely mediated by spin fluctuations. Our results thus provide crucial clues for uncovering the pairing mechanism of iron-based superconductors.

Our results also demonstrate the capability of the combination of ARPES and $in$-$situ$ strain tuning in the determination of pairing symmetry for a superconductor. For example, we could use this method to study how superconductivity interplays with nematicity in BaFe$_2$(As$_{1-x}$P$_x$)$_2$, FeSe and under-doped Ba$_{1-x}$K$_{x}$Fe$_2$As$_2$ where strong nematic fluctuations coexist with superconductivity. It is also very intriguing to study the strain-induced gap modulation in the uniaxially strained K$_{1-x}$Fe$_{2-y}$Se$_2$, Li$_{1-x}$Fe$_x$OHFeSe and other heavily electron-doped FeSe, where the Fermi surface only consists of electron pockets and a nodeless d-wave pairing symmetry has been proposed \cite{ref8,dwave}. Similar methods can also be applied to other complex superconducting materials such as cuprates, Sr$_2$RuO$_4$ and heavy-fermion superconductors. When multiple order parameters intertwine, the uniaxial strain can help to disentangle the strain-sensitive order parameters from the strain-insensitive order parameters. As a supplement to $|\Delta_k|$, the measurement of $\partial|\Delta_k|$ in the perturbed superconducting phase could provide crucial clues for uncovering the pairing mechanisms of various superconducting materials.

This work is supported by the National Natural Science Foundation of China (Grant No. 11888101), National Key Research and Development Program of China (Grant Nos. 2018YFA0305602 and 2016YFA0301003), and National Natural Science Foundation of China (Grant Nos. 91421107 and 11574004).

\end{document}